\documentclass[12pt]{article}
\usepackage{amssymb}
\usepackage{graphicx}
\renewcommand{\theequation}{\arabic{section}.\arabic{equation}}

\begin{document}



\def\a{\alpha}
\def\b{\beta}
\def\d{\delta}
\def\e{\epsilon}
\def\g{\gamma}
\def\h{\mathfrak{h}}
\def\k{\kappa}
\def\l{\lambda}
\def\o{\omega}
\def\p{\wp}
\def\r{\rho}
\def\t{\tau}
\def\s{\sigma}
\def\z{\zeta}
\def\x{\xi}
\def\V={{{\bf\rm{V}}}}
 \def\A{{\cal{A}}}
 \def\B{{\cal{B}}}
 \def\C{{\cal{C}}}
 \def\D{{\cal{D}}}
\def\G{\Gamma}
\def\K{{\cal{K}}}
\def\O{\Omega}
\def\R{\bar{R}}
\def\T{{\cal{T}}}
\def\L{\Lambda}
\def\f{E_{\tau,\eta}(sl_2)}
\def\E{E_{\tau,\eta}(sl_n)}
\def\Zb{\mathbb{Z}}
\def\Cb{\mathbb{C}}

\def\R{\overline{R}}

\newcommand{\be}{\begin{eqnarray}}
\newcommand{\ee}{\end{eqnarray}}
\def\beq{\begin{equation}}
\def\eeq{\end{equation}}
\def\bea{\begin{eqnarray}}
\def\eea{\end{eqnarray}}
\def\ba{\begin{array}}
\def\ea{\end{array}}
\def\no{\nonumber}
\def\le{\langle}
\def\re{\}}
\def\lt{\left}
\def\rt{\right}
\def\non{\nonumber}
\newcommand{\tr}{\mathop{\rm tr}\nolimits}
\newcommand{\sn}{\mathop{\rm sn}\nolimits}
\newcommand{\cn}{\mathop{\rm cn}\nolimits}
\newcommand{\dn}{\mathop{\rm dn}\nolimits}
\newcommand{\kk}{\kappa}
\newcommand{\id}{\mathbb{I}}
\newcommand{\sgn}{\mathop{\rm sgn}\nolimits}
\newcommand{\ch}{\mathop{\rm ch}\nolimits}
\newcommand{\sh}{\mathop{\rm sh}\nolimits}
\newcommand{\tnh}{\mathop{\rm tanh}\nolimits}
\newcommand{\cth}{\mathop{\rm coth}\nolimits}

\newtheorem{Theorem}{Theorem}
\newtheorem{Definition}{Definition}
\newtheorem{Proposition}{Proposition}
\newtheorem{Lemma}{Lemma}
\newtheorem{Corollary}{Corollary}
\newcommand{\proof}[1]{{\bf Proof. }
        #1\begin{flushright}$\Box$\end{flushright}}

\baselineskip=20pt

\newfont{\elevenmib}{cmmib10 scaled\magstep1}
\newcommand{\preprint}{
   \begin{flushleft}
   \end{flushleft}\vspace{-1.3cm}
   \begin{flushright}\normalsize
   \end{flushright}}
\newcommand{\Title}[1]{{\baselineskip=26pt
   \begin{center} \Large \bf #1 \\ \ \\ \end{center}}}
\newcommand{\Author}{\begin{center}
   \large \bf
Junpeng Cao${}^{a, b}$, Shuai Cui${}^a$, Wen-Li Yang${}^{c, d},
\footnote{Corresponding author: wlyang@nwu.edu.cn}$Kangjie
Shi${}^c$ and~Yupeng Wang${}^{a, b} \,\footnote{Corresponding
author: yupeng@iphy.ac.cn}$
 \end{center}}
\newcommand{\Address}{\begin{center}

     ${}^a$Beijing National Laboratory for Condensed Matter
           Physics, Institute of Physics, Chinese Academy of Sciences, Beijing
           100190, China\\
     ${}^b$Collaborative Innovation Center of Quantum Matter, Beijing,
     China\\
     ${}^c$Institute of Modern Physics, Northwest University,
     Xian 710069, China \\
     ${}^d$Beijing Center for Mathematics and Information Interdisciplinary Sciences, Beijing, 100048,  China
   \end{center}}
\newcommand{\Accepted}[1]{\begin{center}
   {\large \sf #1}\\ \vspace{1mm}{\small \sf Accepted for Publication}
   \end{center}}

\preprint \thispagestyle{empty}
\bigskip\bigskip\bigskip

\Title{Exact spectrum of the spin-$s$ Heisenberg chain with generic non-diagonal boundaries } \vspace{1cm}
\Author

\Address \vspace{1cm}

\begin{abstract}

The off-diagonal Bethe ansatz method is generalized to the high spin integrable systems associated with the
$su(2)$ algebra by employing the spin-$s$ isotropic Heisenberg chain model with generic integrable boundaries
as an example. With the fusion techniques, certain closed operator identities for constructing
the functional $T-Q$ relations and the Bethe ansatz equations are derived. It is found that a variety of inhomogeneous $T-Q$
relations obeying the operator product identities can be constructed. Numerical results for
two-site $s=1$ case indicate that an arbitrary choice of the derived $T-Q$ relations is enough to give the complete spectrum of
the transfer matrix.

\vspace{1truecm} \noindent {\it PACS:} 75.10.Pq, 02.30.Ik, 05.30.Jp

\end{abstract}
\newpage

\section{Introduction}
\label{intro} \setcounter{equation}{0}

Among integrable quantum spin chains, the $su(2)$-invariant spin-$s$
Heisenberg chain is particularly interesting due to its relationship
to the Wess-Zumino-Novikov-Witten (WZNW)  models
\cite{wzw1,wzw2,wzw3,ra} and lower dimensional super-symmetric
quantum field theories \cite{Sha78} such as the super-symmetric
sine-Gordon model \cite{Ina95}, the fractional statistics
\cite{fra1} and the multi-channel Kondo problem
\cite{andrei1,andrei2,wang1} when it couples to an impurity spin. The
$s=1$ integrable spin chain model was firstly proposed by Zamalodchikov and Fateev
\cite{zf}. Its generalization to arbitrary $s$ cases was
subsequently constructed via the fusion techniques \cite{fusion}
based on the fundamental $s=\frac12$ representations of the
Yang-Baxter equation \cite{yang,baxter}. Those observations allow
one to diagonalize the models with periodic boundary conditions in
the framework of algebraic Bethe ansatz method (for example, see
\cite{spinsXXX}). On the other hand, the discovery of the boundary
Yang-Baxter equation or the reflection equation \cite{Che84,Sk}
directly stimulated the studies on the exact solutions of the
quantum integrable models with boundary fields. A striking feature
of the reflection equation is that it allows non-diagonal solutions
\cite{de1,GZ}, which leads to the corresponding eigenvalue problem
quite frustrated. Many efforts had been made
\cite{CLSW,Nep04,Yan04-1,Gie05,Gie05-1,Doi06,Baj06,Yan06,YZ3,Gal08,Fra08,Nic12,bell} to approach this nontrivial problem.
However, in a long period of time, the Bethe ansatz solutions
could only be obtained for either constrained boundary parameters
\cite{CLSW} or special crossing parameters \cite{Nep04} associated with spin-$\frac12$ chains or with spin-$s$ chains
\cite{Fra07,mur,barmur,matins}.

Recently, based on the fundamental properties of the $R$-matrix
and the $K$-matrices for quantum integrable models, a systematic
method for solving the eigenvalue problem of integrable models with generic boundary
conditions, i.e., the off-diagonal Bethe ansatz (ODBA) method was
proposed in \cite{cao1} and several long-standing models \cite{cao1,lcysw14,ZCYSW14}  were
then solved. Subsequently, the nested-version of ODBA for the
models associated with $su(n)$ algebra \cite{Cao14}, the application
to the integrable models beyond $A$-type \cite{Hao14} and the
thermodynamic analysis  based on the ODBA solutions \cite{Li14} were developed.
We remark that two other promising methods, namely, the $q$-Onsager algebra method \cite{Bas07} and the separation of variables (SoV)
method \cite{Nic12,nicc14-1} were also  used to approach the spin-$\frac12$ chains with generic integrable boundaries.
Especially, the eigenstate problem for such kind of models with generic inhomogeneity was first approached via the SoV method \cite{Nic12}.
A set of Bethe states for was then constructed in \cite{sigma} and a method for retrieving the Bethe states based on the inhomogeneous $T-Q$ relation \cite{cao1} and the SoV basis \cite{Nic12} was developed in \cite{Cao-14-Bethe-state}. The latter method
allows one to reach the homogeneous limit of the SoV eigenstates \footnote{ It should be emphasized
that the Bethe-type eigenstates of the spin-$\frac{1}{2}$ XXX chain with generic boundaries had challenged for many years and were conjectured in \cite{sigma} and derived in \cite{Cao-14-Bethe-state} very recently after the discovery of the inhomogeneous
$T-Q$ relation in \cite{cao1}. Only together with the
very  inhomogeneous $T-Q$ relation, the SoV state \cite{Nic12} might be transformed into a Bethe state which possesses a well-defined homogeneous limit.}
and provides a clear connection
 among the SoV approach,
the algebraic Bethe Ansatz and the ODBA.

The high spin models with periodic \cite{zf,fusion,spinsXXX} and  diagonal  \cite{MNR,FLSU,Do2} boundaries have been
extensively studied. Even the
most general integrable boundary condition (corresponding to the
non-diagonal reflection matrix) for the spin-$1$ model has been
known for many years \cite{Ina96}, the exact solutions of the models
with non-diagonal boundaries were known only for some special cases
such as the boundary parameters obeying some constraint \cite{Fra07}
or  the crossing parameter taking some special value (e.g., roots of
unity) \cite{mur,barmur}.
In this paper, we show that the ODBA
method can also be applied to the $su(2)$-invariant spin-$s$ chain
with generic crossing parameter and generic integrable boundaries
\footnote{A hierarchy procedure for the isotropic open chain constructed with higher
dimensional auxiliary spaces
and each of its $N$ quantum spaces are all spin-$\frac{1}{2}$ (i.e., two-dimensional)
 was proposed in \cite{Nep13-1}.}.  The
outline of the paper is the following: Section 2 serves as an
introduction to some notations and the fusion
procedure. In Section 3, after briefly
reviewing the fusion hierarchy \cite{Fra07} of the high spin
transfer matrices we derive certain closed  operator product identities for
the fundamental spin-$(\frac{1}{2},s)$ transfer matrix by using some
intrinsic properties of the high spin  $R$-matrix ($R^{(s,s)}(u)$)
and $K$-matrices ($K^{\pm(s)}(u)$). The asymptotic behavior of the
transfer matrix is also obtained. Section 4 is devoted to the
construction of the inhomogeneous $T-Q$ relations and the
corresponding Bethe ansatz equations (BAEs). Taking the spin-$1$ XXX
chain as an example, we present numerical results for the model with
some small number of sites, which indicate that an arbitrary choice of the
derived $T-Q$ relations is enough to give the complete set of spectrum of the
transfer matrix. In section 5, we summarize our results and give
some discussions. In Appendix A, we prove that each solution of our functional equations can be parameterized in terms of
a variety of inhomogeneous $T-Q$ relations and therefore that different $T-Q$ relations only indicate different
parameterizations but not new solutions.

\section{Transfer matrices for the spin-$s$ XXX spin chain}
\label{aniso} \setcounter{equation}{0}

\subsection{Fusion of the $R$-matrices and the $K$-matrices}
Throughout,  $V_i$ denotes  a $(2l_i+1)$-dimensional linear space
($\Cb^{2l_i+1}$) which endows an irreducible representation of
$su(2)$ algebra with spin $l_i$. The $R$-matrix  $R^{(l_i,l_j)}_{ij}(u)$,
denoted as the spin-$(l_i,l_j)$ $R$-matrix, is a linear operator
acting in $V_i\otimes V_j$. The $R$-matrix satisfies the following
quantum Yang-Baxter equation (QYBE) \cite{yang,baxter} \bea
R_{12}^{(l_1,l_2)}(u-v) R_{13}^{(l_1,l_3)}(u)R_{23}^{(l_2,l_3)}(v)=
R_{23}^{(l_2,l_3)}(v)
R_{13}^{(l_1,l_3)}(u)R_{12}^{(l_1,l_2)}(u-v).\label{QYBE} \eea Here
and below we adopt the standard notations: for any matrix $A\in {\rm
End}({ V})$, $A_j$ is an embedding operator in the tensor
space ${ V}\otimes { V}\otimes\cdots$, which acts as $A$ on the
$j$-th space and as identity on the other factor spaces; $R_{ij}(u)$
is an embedding operator of $R$-matrix in the tensor space, which
acts as identity on the factor spaces except for the $i$-th and
$j$-th ones.

The fundamental spin-$(\frac{1}{2},s)$ $R$-matrix
$R^{(\frac{1}{2},s)}_{12}(u)$ defined in spin-$\frac{1}{2}$ (i.e.,
two-dimensional) auxiliary space and spin-$s$ (i.e.,
$(2s+1)$-dimensional) quantum space is given by \cite{fusion} \bea
R^{(\frac{1}{2},s)}_{12}(u)=u+\frac{\eta}{2}+\eta\,
\vec{\s}_1\cdot \vec{S}_2,\label{R-1s} \eea
where $\eta$ is the crossing parameter, $\vec{\s}$ are
the Pauli matrices and $\vec{S}$ are the spin-$s$ realization of the $su(2)$
generators. For the simplest case, i.e.,
$s=\frac{1}{2}$ case the corresponding $R$-matrix reads

\bea
R^{(\frac{1}{2},\frac{1}{2})}(u) = \left(
\begin{array}{cccc}
    u + \eta &0            &0           &0            \\
    0                 &u     & \eta  &0            \\
    0                 &\eta   &  u    &0            \\
    0                 &0            &0           &u + \eta
\end{array} \right).
\label{R-matrix11}
\eea

\noindent Besides the QYBE (\ref{QYBE}), the $R$-matrix
(\ref{R-matrix11}) also enjoys the following properties, \bea
&&\hspace{-1.5cm}\mbox{Initial
condition}:\,R^{(\frac{1}{2},\frac{1}{2})}_{12}(0)=\eta P_{12},\label{Int-R}\\[6pt]
&&\hspace{-1.5cm}\mbox{Unitary
relation}:\,R^{(\frac{1}{2},\frac{1}{2})}_{12}(u)R^{(\frac{1}{2},\frac{1}{2})}_{21}(-u)= -\xi(u)\,{\rm id},
\quad \xi(u)=(u+\eta)(u-\eta),\label{Unitarity}\\[6pt]
&&\hspace{-1.5cm}\mbox{Crossing
relation}:\,R^{(\frac{1}{2},\frac{1}{2})}_{12}(u)=V_1\{R^{(\frac{1}{2},\frac{1}{2})}_{12}\}^{t_2}(-u-\eta)V_1,\quad
V=-i\s^y,
\label{crosing-unitarity}\\[6pt]
&&\hspace{-1.5cm}\mbox{PT-symmetry}:\,R^{(\frac{1}{2},\frac{1}{2})}_{12}(u)=R^{(\frac{1}{2},\frac{1}{2})}_{21}(u)
=\{R^{(\frac{1}{2},\frac{1}{2})}\}^{t_1\,t_2}_{12}(u),\label{PT}\\[6pt]
&&\hspace{-1.5cm}\mbox{Fusion conditions}:\,R^{(\frac{1}{2},\frac{1}{2})}_{12}(\pm\eta)=\eta(\pm1+ P_{12})=\pm 2\eta P_{12}^{\pm}.\label{Fusion-Con}
\eea
Here $R^{(\frac{1}{2},\frac{1}{2})}_{21}(u)=P_{12}R^{(\frac{1}{2},\frac{1}{2})}_{12}(u)P_{12}$ with $P_{12}$ being
the permutation operator and $t_i$ denotes transposition
in the $i$-th space. Using the fusion procedure \cite{fusion} the spin-$(\frac{1}{2},s)$ $R$-matrix $R^{(\frac{1}{2},s)}_{12}(u)$
can be obtained by the symmetric fusion of the spin-$(\frac{1}{2},\frac{1}{2})$ $R$-matrix\footnote{It is worth noting that, strictly speaking,
after a similarity transformation the fused $R$-matrices (\ref{fusedR0103}) and (\ref{fusedR010013})
and the fused $K$-matrices (see below (\ref{fusedKmatrix0103}) and (\ref{Correspond}))
all contain null rows and columns. Once these rows and columns are removed, the matrices have
the correct size.}
\be R^{({1\over 2},s)}_{a \{ 1 \cdots 2s \}
}(u) = \frac{1}{\prod_{k=1}^{2s-1}(u\hspace{-0.04truecm}+\hspace{-0.04truecm}
(\frac{1}{2}\hspace{-0.04truecm}-\hspace{-0.04truecm}s\hspace{-0.04truecm}+\hspace{-0.04truecm}k)\eta)}
P^{+}_{\{1 \cdots 2s\}}\prod_{k=1}^{2s} \lt\{R^{(\frac{1}{2},\frac{1}{2})}_{a,
k}(u\hspace{-0.04truecm}+\hspace{-0.04truecm}(k\hspace{-0.04truecm}-\hspace{-0.04truecm}\frac{1}{2}
\hspace{-0.04truecm}-\hspace{-0.04truecm}s)\eta)\rt\} P^{+}_{\{1
\cdots 2s\}}, \label{fusedR0103}
\ee where $P^{+}_{\{1 \cdots 2s\}}$ is the symmetric projector given by
\bea
P^{+}_{1,\cdots,2s}=\frac{1}{(2s)!}\prod_{k=1}^{2s}\lt(\sum_{l=1}^k P_{l\,k}\rt).\label{Symmetric-P}
\eea
Similarly, from the spin-$({1\over 2}, s)$ $R$-matrix we can also extend the
auxiliary space from $\frac{1}{2}$ to $j$ to obtain the
spin-$(j, s)$ $R$-matrix by the symmetric fusion
\be && R^{(j,s)}_{\{1 \cdots
2j\} \{ 1 \cdots 2s \} }(u) = P^{+}_{\{1 \cdots 2j\}}
\prod_{k=1}^{2j}\lt\{R^{(\frac{1}{2},s)}_{k, \{ 1 \cdots 2s \}}(u+(k-j-\frac{1}{2})\eta)\rt\}
 P^{+}_{\{1 \cdots 2j\}}. \label{fusedR010013}
\ee We remark that the $R$-matrices in the products (\ref{fusedR010013}) and (\ref{fusedR0103}) are
in the order of increasing $k$. One can demonstrate that the fused $R$-matrices (\ref{R-1s}) and (\ref{fusedR0103}) also satisfy the associated
QYBE (\ref{QYBE}) with the help of (\ref{Fusion-Con}). Direct calculation  shows
that the spin-$(s,s)$ $R$-matrix can be given by \cite{fusion}
\bea
R^{(s,s)}=\prod_{j=1}^{2s}(u-j\eta)\,\sum_{l=0}^{2s}\prod_{k=1}^l\frac{u+k\eta}{u-k\eta}\,\mathbf{P}^{(l)},\label{R-ss}
\eea where $\mathbf{P}^{(l)}$ is a projector acting on the tensor product of two spin-$s$ spaces and projects the tensor
space into the irreducible subspace of spin-$l$ (i.e., $(2l+1)$-dimensional subspace). In particular, the fundamental
spin-$(\frac{1}{2},s)$ and the fused spin-$(s,s)$ $R$-matrix
possess the following important properties
\bea
&&\hspace{-1.5cm}
\mbox{Unitary
relation}:\,R^{(\frac{1}{2},s)}_{12}(u)R^{(s,\frac{1}{2})}_{21}(-u)= -(u+(\frac{1}{2}+s)\eta)(u-(\frac{1}{2}+s)\eta)\,{\rm id},\label{Unitarity-ss}\\[2pt]
&&\hspace{-1.5cm}\mbox{ Initial
condition}:\,R^{(s,s)}_{12}(0)= \eta^{2s}(2s)!P_{12},\label{Int-R-ss}\\[2pt]
&&\hspace{-1.5cm}\mbox{ Fusion condition}:\,R^{(s,s)}_{12}(-\eta)=(-1)^{2s}\eta^{2s}(2s+1)!\,\mathbf{P}^{(0)}. \label{Fusion-Con-1}
\eea The projector $\mathbf{P}^{(0)}$ projects the tensor product of two spin-$s$ spaces to the singlet space, namely,
\bea
\mathbf{P}^{(0)}=|\Phi_0\rangle\langle\Phi_0|,\quad |\Phi_0\rangle=\frac{1}{\sqrt{2s+1}}\sum_{l=0}^{2s}(-1)^l|s-l\rangle\otimes|-s+l\rangle,\label{Singlet}
\eea where $\{|m\rangle,\,m=s,s-1,\ldots,-s\}$ spans the spin-$s$ representation of $su(2)$ algebra and forms an orthonormal basis of it. The very properties
(\ref{Unitarity-ss}), (\ref{Int-R-ss}) and (\ref{Fusion-Con-1}) are the analogues of (\ref{Unitarity}), (\ref{Int-R}) and (\ref{Fusion-Con}) for the high spin case.

Having defined the fused-$R$ matrices, one can analogously construct the fused-$K$ matrices by using the methods developed in \cite{MNR,fusion2}
as follows.  The fused $K^-$ matrices (e.g the spin-$j$ $K^-$ matrix) is given by
\be K^{- (j)}_{\{a\}}(u) &=& P_{\{a\}}^{+}
\prod_{k=1}^{2j} \Bigg\{ \left[ \prod_{l=1}^{k-1}
R^{(\frac{1}{2},\frac{1}{2})}_{a_{l}a_{k}}
(2u+(k+l-2j-1)\eta) \right] \non \\[6pt]
&&\times  K^{- (\frac{1}{2})}_{a_{k}}(u+(k-j-\frac{1}{2})\eta)
\Bigg\} P_{\{a\}}^{+}. \label{fusedKmatrix0103} \ee
In this paper we adopt the most general non-diagonal spin-$\frac{1}{2}$ $K$-matrix
$K^{- (\frac{1}{2})}(u)$ \cite{de1, GZ}
\begin{eqnarray}
K^{-(\frac12)}(u)=\left(\begin{array}{cc} p_- + u & \alpha_- u \\
\alpha_- u & p_- -u
\end{array}\right), \label{K-}
\end{eqnarray}
where $p_-$ and $\alpha_-$ are some boundary parameters. It is noted that the products  of braces $\{\ldots\}$ in (\ref{fusedKmatrix0103})
are in the order of increasing $k$. The fused $K^{- (j)}_{\{a\}}(u)$ matrices satisfy
the following reflection equation \cite{Che84,Fra07}
\be \lefteqn{R^{(j,s)}_{\{a\}
\{b\}}(u-v)\, K^{- (j)}_{\{a\}}(u)\,
R^{(s,j)}_{\{b\} \{a\}}(u+v)\, K^{- (s)}_{\{b\}}(v)}\non \\[6pt]
& & =K^{- (s)}_{\{b\}}(v)\, R^{(j,s)}_{\{a\} \{b\}}(u+v)\, K^{-
(j)}_{\{a\}}(u)\, R^{(s,j)}_{\{b\} \{a\}}(u-v) \,.\label{RE} \ee
The fused dual reflection matrices $K^{+ (j) }$ \cite{Sk} are given by \be
K^{+ (j)}_{\{a\}}(u) = {1\over f^{(j)}(u)}\,K^{- (j)}_{\{a\}}
(-u-\eta)\Big\vert_{(p_-,\alpha_-)\rightarrow
(p_+,-\alpha_+)} \,, \label{Correspond}\ee with
\be f^{(j)}(u) = \prod_{l=1}^{2j-1}\prod_{k=1}^{l} [-\xi(
2u + (l+k+1-2j)\eta) ]. \label{Kplusnormalization0103} \ee
Particularly, the fundamental one
$K^{+(\frac{1}{2})}(u)$ is
\begin{eqnarray}
K^{+(\frac12)}(u)=\left(\begin{array}{cc} p_+ -u-\eta & \alpha_+ (u+\eta) \\
\alpha_+ (u+\eta) & p_+ +u+\eta
\end{array}\right)=K^{- (\frac12)}(-u-\eta)\Big\vert_{(p_-,
\alpha_-)\rightarrow (p_+, -\alpha_+)},\ \label{K-0102}
\end{eqnarray}
where $p_+$ and $\alpha_+$ are some boundary parameters.

\subsection{Fused transfer matrices}
In a similar way to that developed by Sklyanin \cite{Sk} for the spin-$\frac12$ case, one can construct a transfer matrix $t^{(j,s)}(u)$ whose auxiliary space is spin-$j$ ($(2j + 1)$-
dimensional) and each of its $N$ quantum spaces are spin-$s$ ($(2s + 1)$-dimensional) following the method in \cite{Fra07},
for any $j, s \in \{\frac{1}{2}, 1,
\frac{3}{2}, \ldots\}$. The fused
transfer matrix  $t^{(j,s)}(u)$ can be constructed by the fused $R$-matrices and $K$-matrices as follows \cite{Sk,Fra07}
\be t^{(j,s)}(u) = \tr_{\{a\}}
K^{+ (j)}_{\{a\}}(u)\, T^{(j,s)}_{\{a\}}(u)\, K^{- (j)}_{\{a\}}(u)\,
\hat T^{(j,s)}_{\{a\}}(u) \,, \label{Fused-transfer}
\ee where $T^{(j,s)}_{\{a\}}(u)$ and
$\hat T^{(j,s)}_{\{a\}}(u)$ are the fused one-row monodromy matrices given by
\be T^{(j,s)}_{\{a\}}(u) &=& R^{(j,s)}_{\{a\},
\{b^{[N]}\}}(u-\theta_N) \ldots
R^{(j,s)}_{\{a\}, \{b^{[1]}\}}(u-\theta_1) \,, \non \\[6pt]
\hat T^{(j,s)}_{\{a\}}(u) &=& R^{(s,j)}_{
\{b^{[1]}\},\{a\}}(u+\theta_1) \ldots R^{(s,j)}_{
\{b^{[N]}\},\{a\}}(u+\theta_N) \,. \ee Here $\{\theta_j|j=1,\ldots, N\}$ are arbitrary free complex parameters which are usually
called the inhomogeneous parameters. The QYBE (\ref{QYBE}), the reflection equation (\ref{RE}) and its dual version lead to that
these transfer matrices with different spectral parameters are mutually commutative for
arbitrary $j , j',s \in
\{\frac{1}{2}, 1, \frac{3}{2}, \ldots \}$\be \left[ t^{(j,s)}(u)
\,, t^{(j',s)}(v) \right] = 0 \,. \label{commutativity0103} \ee
Therefore $t^{(j,s)}(u)$ serve as the generating functionals of the conserved quantities.

\section{Fusion hierarchy and operator identities} \label{anisoopen} \setcounter{equation}{0}

\subsection{Operator identities}
Let us fix an $s\in \{\frac{1}{2},1,\frac{3}{2},\ldots\}$, i.e.,
each of the $N$ quantum spaces is described by a spin $\vec S$ ($(2s + 1)$-dimensional).
The fused transfer matrices $\{t^{(j,s)}(u)\}$ given by (\ref{Fused-transfer}) obey the following fusion hierarchy relation \cite{MNR,
fusion2,Fra07}
\be t^{(\frac{1}{2},s)}(u)\, t^{(j-\frac{1}{2},s)}(u- j\eta)&=&
t^{(j,s)}(u-(j-\frac{1}{2})\eta) + \delta^{(s)}(u)\,
t^{(j-1,s)}(u-(j+\frac{1}{2})\eta),\no\\[6pt]
j &=&\frac
12, 1, \frac{3}{2}, \cdots, \label{hierarchy0103}
\ee
where we have used the convention $t^{(0,s)}={\rm id}$. The coefficient function
$\delta^{(s)}(u)$ related to the quantum determinant is given by
\be
\delta^{(s)}(u) &=&\frac{(2u-2\eta)(2u+2\eta)}{(2u-\eta)(2u+\eta)}
((1+\alpha_-^2)u^2-p_-^2)((1+\alpha_+^2)u^2-p_+^2)\no\\[2pt]
&&\quad \times \prod_{l=1}^N (u-\theta_l+(\frac{1}{2}+s)\eta)(u+\theta_l+(\frac{1}{2}+s)\eta)\no\\[2pt]
&&\quad \times \prod_{l=1}^N (u-\theta_l-(\frac{1}{2}+s)\eta)(u+\theta_l-(\frac{1}{2}+s)\eta).\label{delta0103}
\ee Using the recursive relation (\ref{hierarchy0103}), we can express the fused transfer matrix $t^{(j,s)}(u)$ in
terms of the fundamental one $t^{(\frac{1}{2},s)}(u)$ with a $2j$-order functional relation as follows:
\bea
 t^{(j,s)}(u)&=&t^{(\frac{1}{2},s)}(u+(j-\frac{1}{2})\eta)\,t^{(\frac{1}{2},s)}(u+(j-\frac{1}{2})\eta-\eta)\ldots
                t^{(\frac{1}{2},s)}(u-(j-\frac{1}{2})\eta)\no\\[2pt]
             &&-\d^{(s)}(u+(j-\frac{1}{2})\eta)\,t^{(\frac{1}{2},s)}(u+(j-\frac{1}{2})\eta-2\eta)\ldots
             \,t^{(\frac{1}{2},s)}(u-(j-\frac{1}{2})\eta)\no\\[2pt]
             &&-\d^{(s)}(u+(j-\frac{1}{2})\eta-\eta)\,t^{(\frac{1}{2},s)}(u+(j-\frac{1}{2})\eta)\,
             t^{(\frac{1}{2},s)}(u+(j-\frac{1}{2})\eta-3\eta)\no\\[2pt]
             &&\quad\quad \times\ldots
             \,t^{(\frac{1}{2},s)}(u-(j-\frac{1}{2})\eta)\no\\[2pt]
             &&\vdots\no\\
             &&-\d^{(s)}(u\hspace{-0.04truecm}-\hspace{-0.04truecm}(j\hspace{-0.04truecm}-\hspace{-0.04truecm}\frac{1}{2})\eta
             \hspace{-0.04truecm}+\hspace{-0.04truecm}\eta)\,t^{(\frac{1}{2},s)}
             (u\hspace{-0.04truecm}+\hspace{-0.04truecm}(j\hspace{-0.04truecm}-\hspace{-0.04truecm}\frac{1}{2})\eta)\ldots
             \,t^{(\frac{1}{2},s)}(u\hspace{-0.04truecm}-\hspace{-0.04truecm}
             (j\hspace{-0.04truecm}-\hspace{-0.04truecm}\frac{1}{2})\eta\hspace{-0.04truecm}+\hspace{-0.04truecm}2\eta)\no\\[2pt]
             &&+\ldots.\label{Fusion-Hier}
\eea
For example, the first three fused transfer matrices are given by
\bea
  t^{(1,s)}(u)&=& t^{(\frac{1}{2},s)}(u+\frac{\eta}{2})\,t^{(\frac{1}{2},s)}(u-\frac{\eta}{2})-\d^{(s)}(u+\frac{\eta}{2}),\label{Fused-T-1}\\[2pt]
  t^{(\frac{3}{2},s)}(u)&=& t^{(\frac{1}{2},s)}(u+\eta)\,t^{(\frac{1}{2},s)}(u)\,t^{(\frac{1}{2},s)}(u-\eta)
                           -\d^{(s)}(u+\eta)\,t^{(\frac{1}{2},s)}(u-\eta)\no\\[2pt]
                        && -\d^{(s)}(u)\,t^{(\frac{1}{2},s)}(u+\eta),\label{Fused-T-2}\\[2pt]
  t^{(2,s)}(u)&=& t^{(\frac{1}{2},s)}(u+\frac{3\eta}{2})\,t^{(\frac{1}{2},s)}(u+\frac{\eta}{2})
                   \,t^{(\frac{1}{2},s)}(u-\frac{\eta}{2})\,t^{(\frac{1}{2},s)}(u-\frac{3\eta}{2})\no\\[2pt]
              &&-\d^{(s)}(u+\frac{3\eta}{2})\,t^{(\frac{1}{2},s)}(u-\frac{\eta}{2})\,t^{(\frac{1}{2},s)}(u-\frac{3\eta}{2})\no\\[2pt]
              &&-\d^{(s)}(u+\frac{\eta}{2})\,t^{(\frac{1}{2},s)}(u+\frac{3\eta}{2})\,t^{(\frac{1}{2},s)}(u-\frac{3\eta}{2})\no\\[2pt]
              &&-\d^{(s)}(u-\frac{\eta}{2})\,t^{(\frac{1}{2},s)}(u+\frac{3\eta}{2})\,t^{(\frac{1}{2},s)}(u+\frac{\eta}{2})\no\\[2pt]
              &&+\d^{(s)}(u+\frac{3\eta}{2})\,\d^{(s)}(u-\frac{\eta}{2}).\label{Fused-T-3}
\eea
Keeping the very properties (\ref{Unitarity-ss})-(\ref{Fusion-Con-1}) in mind and following the method developed  in \cite{Cao14,Hao14}, after
a tedious calculation, we find that the spin-$(s,s)$ transfer matrix satisfies the following operator identities,
\footnote{Alternatively, one can show that there exist some
operator identities between $t^{(s,s)}(u)$ and  $t^{(s,s)}(u)$ at some special points.
These relations are equivalent to (\ref{Operator-identity-1}) in the sense that they give rise to the same inhomogeneous $T-Q$
relation (see below (\ref{T-Q-ansatz-1})).}
\bea
t^{(s,s)}(\theta_j) \,t^{(\frac{1}{2},s)}(\theta_j-(\frac{1}{2}+s)\eta)=\d^{(s)}(\theta_j+(\frac{1}{2}-s)\eta)\,
t^{(s-\frac{1}{2},s)}(\theta_j+\frac{\eta}{2}),\quad j=1,\ldots,N.\label{Operator-identity-1}
\eea

The $R$-matrix (\ref{R-1s}) and the $K$-matrices (\ref{K-}) and (\ref{K-0102})
imply that the transfer matrix $t^{(\frac{1}{2},s)}(u)$ possesses the following properties:
\bea
&&t^{(\frac{1}{2},s)}(0)=2p_-p_+\prod_{l=1}^N(\theta_l+(\frac{1}{2}+s)\eta)(-\theta_l+(\frac{1}{2}+s)\eta)\times {\rm id},
 \label{Operator-Int-1}\\[2pt]
&&t^{(\frac{1}{2},s)}(u)\lt|_{u\rightarrow\infty}\rt.=2(\a_-\a_+-1)u^{2N+2}\times {\rm id}+\ldots,\label{Operator-Asy}\\[2pt]
&&t^{(\frac{1}{2},s)}(-u-\eta)=t^{(\frac{1}{2},s)}(u).\label{Operator-crossing}
\eea

The analyticities of the spin-$(\frac{1}{2},s)$ $R$-matrix and spin-$\frac{1}{2}$ $K$-matrices and the
property (\ref{Operator-Asy}) imply that the transfer matrix $t^{(\frac{1}{2},s)}(u)$, as a function of $u$,
is a polynomial of degree $2N+2$. The fusion hierarchy relation (\ref{hierarchy0103}) gives rise to that all the other fused transfer
matrix  $t^{(j,s)}(u)$ can be
expressed in terms of $t^{(\frac{1}{2},s)}(u)$ (see (\ref{Fusion-Hier})). Therefore, the very operator
identities (\ref{Operator-identity-1}) lead to $N$ constraints on the fundamental transfer matrix
$t^{(\frac{1}{2},s)}(u)$. Thus the relations (\ref{Operator-identity-1}) and (\ref{Operator-Int-1})-(\ref{Operator-crossing})
are believed to completely  characterize the eigenvalues of the fundamental transfer matrix $t^{(\frac{1}{2},s)}(u)$ (as a consequence, also
determine the eigenvalues of all the transfer matrices $\{t^{(j,s)}(u)\}$).

\subsection{Functional relations of the eigenvalues}
The commutativity (\ref{commutativity0103}) of the fused transfer matrices $\{t^{(j,s)}(u)\}$ with different spectral
parameters implies that they have common eigenstates. Let $|\Psi\rangle$  be a common eigenstate of these
fused transfer matrices with the eigenvalues $\Lambda^{(j,s)}(u)$
\bea
t^{(j,s)}(u)|\Psi\rangle =\Lambda^{(j,s)}(u)|\Psi\rangle.
\eea The fusion hierarchy relation (\ref{hierarchy0103}) of the fused transfer matrices allows one to express all the eigenvalues
$\Lambda^{(j,s)}(u)$ in terms of the fundamental one $\Lambda^{(\frac{1}{2},s)}(u)$ by the following recursive relations
\be \L^{(\frac{1}{2},s)}(u)\, \L^{(j\hspace{-0.04truecm}-\hspace{-0.04truecm}\frac{1}{2},s)}
(u\hspace{-0.04truecm}-\hspace{-0.04truecm} j\eta)&=&\hspace{-0.08truecm}
\L^{(j,s)}(u\hspace{-0.04truecm}-\hspace{-0.04truecm}
(j\hspace{-0.04truecm}-\hspace{-0.04truecm}\frac{1}{2})\eta) + \delta^{(s)}(u)\,
\L^{(j-1,s)}(u\hspace{-0.04truecm}-\hspace{-0.04truecm}(j\hspace{-0.04truecm}+\hspace{-0.04truecm}\frac{1}{2})\eta),\no\\
j &=&\frac{1}{2}, 1, \frac{3}{2}, \cdots. \label{Eigenvalue-hier}
\ee  Here $\L^{(0,s)}(u)=1$ and the coefficient function $\d^{(s)}(u)$ is given by (\ref{delta0103}). The very operator identities
(\ref{Operator-identity-1}) imply that the  eigenvalue
$\Lambda^{(s,s)}(u)$  satisfies the same relations
\footnote{It should be emphasized that the operator identities (\ref{Operator-identity-1}) are
stronger than the functional relations (\ref{Eigenvalue-identity}) due to the fact that in some extreme case, the transfer matrix cannot be
diagonalized (i.e., the transfer matrix has non-trivial Jordan blocks \cite{Kor08}) and one cannot derive the operator identities (\ref{Operator-identity-1})
only from its eigenvalue version (\ref{Eigenvalue-identity}).}
\bea
\L^{(s,s)}(\theta_j) \,\L^{(\frac{1}{2},s)}(\theta_j\hspace{-0.04truecm}-\hspace{-0.04truecm}
(\frac{1}{2}\hspace{-0.04truecm}+\hspace{-0.04truecm}s)\eta)=\d^{(s)}(\theta_j\hspace{-0.04truecm}+\hspace{-0.04truecm}
(\frac{1}{2}\hspace{-0.04truecm}-\hspace{-0.04truecm}s)\eta)\,
\L^{(s-\frac{1}{2},s)}(\theta_j\hspace{-0.04truecm}+\hspace{-0.04truecm}\frac{\eta}{2}),\quad j=1,\ldots,N.\label{Eigenvalue-identity}
\eea The properties of the transfer matrix $t^{(\frac{1}{2},s)}(u)$
given by (\ref{Operator-Int-1})-(\ref{Operator-crossing}) give rise to that the corresponding eigenvalue $\Lambda^{(\frac{1}{2},s)}(u)$
satisfies the following relations
\bea
&&\L^{(\frac{1}{2},s)}(0)=2p_-p_+\prod_{l=1}^N(\theta_l+(\frac{1}{2}+s)\eta)(-\theta_l+(\frac{1}{2}+s)\eta),
 \label{Eigenvalue-Int-1}\\[2pt]
&&\L^{(\frac{1}{2},s)}(u)\lt|_{u\rightarrow\infty}\rt.=2(\a_-\a_+-1)u^{2N+2}+\ldots,\label{Eigenvalue-Asy}\\[2pt]
&&\L^{(\frac{1}{2},s)}(-u-\eta)=\L^{(\frac{1}{2},s)}(u).\label{Eigenvalue-crossing}
\eea The analyticities of the spin-$(\frac{1}{2},s)$ $R$-matrix and
spin-$\frac{1}{2}$ $K$-matrices and the property
(\ref{Eigenvalue-Asy}) imply that the eigenvalue
$\Lambda^{(\frac{1}{2},s)}(u)$ possesses the following analytical
property \bea \L^{(\frac{1}{2},s)}(u) \mbox{, as a function of $u$,
is a polynomial of degree $2N+2$}.\label{Eigenvalue-Anal} \eea
Namely, $\L^{(\frac{1}{2},s)}(u)$ is a polynomial of $u$ with $2N+3$
unknown coefficients. The crossing relation
(\ref{Eigenvalue-crossing}) reduces the number of the independent
unknown coefficients to $N+2$. Therefore the  relations
(\ref{Eigenvalue-hier})-(\ref{Eigenvalue-Anal}) are believed to completely
characterize  the spectrum of the fundamental spin-$(\frac{1}{2},s)$
transfer matrix $t^{(\frac{1}{2},s)}(u)$.

For the $s=\frac{1}{2}$ case, the relations (\ref{Eigenvalue-hier})-(\ref{Eigenvalue-Anal}) are reduced to those used in \cite{cao1} to determine the spectrum of the corresponding transfer matrix. The eigenstates associated with
each solution of the resulting  relations were constructed in \cite{Nic12} in the framework of the SoV method. In such a sense, each solution corresponds to a correct eigenvalue of the transfer matrix. Since all the eigenvalues of the transfer matrix belong to the solution set of (\ref{Eigenvalue-hier})-(\ref{Eigenvalue-Anal}), we conclude that in the spin-$\frac{1}{2}$ case our functional relations characterize the spectrum completely. It is remarked that
the corresponding Bethe states were given in \cite{sigma,Cao-14-Bethe-state}. These Bethe states have well-defined
homogeneous limits and allows one to study the corresponding homogeneous open chain directly. For the spin-$1$ case,
the numerical results in subsection 4.2 for $N=2$ case also suggest that  the equations
(\ref{Eigenvalue-hier})-(\ref{Eigenvalue-Anal}) indeed give the complete spectrum of the transfer
matrix.

\section{$T-Q$ relation} \label{ansatz} \setcounter{equation}{0}

\subsection{ Eigenvalues of the fundamental transfer matrix} \label{T-QRo}
Following the method developed in \cite{cao1}, let us introduce the following inhomogeneous
$T-Q$ relation
\bea
 \L^{(\frac{1}{2},s)}(u)&=&a^{(s)}(u)\frac{Q(u-\eta)\,Q_1(u-\eta)}{Q(u)\,Q_2(u)}
                           +d^{(s)}(u)\frac{Q(u+\eta)\,Q_2(u+\eta)}{Q(u)\,Q_1(u)}\no\\[6pt]
 &&+c\,u(u+\eta)\frac{(u(u+\eta))^mF^{(s)}(u)}{Q(u)\,Q_1(u)\,Q_2(u)},\label{T-Q-ansatz-1}
\eea where $m$ is a non-negative integer and the functions
$a^{(s)}(u)$, $d^{(s)}(u)$, $F^{(s)}(u)$ and the constant $c$ are
given by \bea
a^{(s)}(u)&=&\frac{2u+2\eta}{2u+\eta}(\sqrt{1+\a_+^2}\,u+p_+)(\sqrt{1+\a_-^2}\,u+p_-) \no\\[2pt]
          &&\quad \times \prod_{l=1}^N(u-\theta_l+(\frac{1}{2}+s)\eta)(u+\theta_l+(\frac{1}{2}+s)\eta),\label{a-function}\\[2pt]
d^{(s)}(u)&=&a^{(s)}(-u-\eta),\label{d-function}\\[2pt]
F^{(s)}(u)&=&\prod_{l=1}^N\prod_{k=0}^{2s}(u-\theta_l+(\frac{1}{2}-s+k)\eta)(u+\theta_l+(\frac{1}{2}-s+k)\eta),\label{F-function}\\[2pt]
c&=&2(\a_-\a_+-1-\sqrt{(1+\a^2_-)(1+\a^2_+)}).\label{c-constant}
\eea The $Q_i(u)$ functions are parameterized by $2sN+m$ parameters
$\{\l_j|j=1,\ldots,2sN+m-2M\}$ and $\{\mu_j|j=1,\ldots,2M\}$ ($M$
being a non-negative integer) as \footnote{One can easily check that the zero points of any $Q_i(u)$ must not take the values of $\{\theta_j+(\frac{1}{2}-s+k)\eta|k=0,\ldots,2s,\,j=1,\ldots, N\}$ and their crossing points
($\{-\theta_j-(\frac{1}{2}-s+k)\eta-\eta|k=0,\ldots,2s,\,j=1,\ldots, N\}$). Otherwise $\L^{(\frac{1}{2},s)}(u)$ given by (\ref {T-Q-ansatz-1}) does not satisfy (\ref{Eigenvalue-identity}).}
\bea
Q(u)&=&\prod_{j=1}^{2sN+m-2M}(u-\l_j)(u+\l_j+\eta)=Q(-u-\eta),\label{Q-function}\\[2pt]
Q_1(u)&=&\prod_{j=1}^{2M}(u-\mu_j)=Q_2(-u-\eta),\label{Q1-function}\\[2pt]
Q_2(u)&=&\prod_{j=1}^{2M}(u+\mu_j+\eta)=Q_1(-u-\eta).\label{Q2-function}
\eea One can   check that the $T-Q$ relation (\ref{T-Q-ansatz-1}) does satisfy the relations
(\ref{Eigenvalue-Int-1})-(\ref{Eigenvalue-crossing}). The explicit expression (\ref{F-function}) of the function
$F^{(s)}(u)$ implies that
\bea
F^{(s)}(\theta_j+(s-\frac{1}{2}-k)\eta))=0,\quad {\rm for}\,\, k=0,1,\ldots,2s,\quad j=1,\ldots,N.\no
\eea  Combining the above equations and the fusion hierarchy relations (\ref{Eigenvalue-hier}), we can evaluate
 $\L^{(s,s)}(u)$, $\L^{(s-\frac{1}{2},s)}(u)$ and $\L^{(\frac{1}{2},s)}(u)$   at the points
$\theta_j$, $\theta_j+\frac{\eta}{2}$ and $\theta_j-(\frac{1}{2}+s)\eta$ respectively
\bea
\hspace{-0.8truecm}\L^{(s,s)}(\theta_j)\hspace{-0.28truecm}&=& \hspace{-0.28truecm}\frac{Q(\theta_j\hspace{-0.04truecm}-\hspace{-0.04truecm}
                    (s\hspace{-0.04truecm}+\hspace{-0.04truecm}\frac{1}{2})\eta)}
                    {Q(\theta_j\hspace{-0.04truecm}+\hspace{-0.04truecm}
                    (s\hspace{-0.04truecm}-\hspace{-0.04truecm}\frac{1}{2})\eta)}\hspace{-0.08truecm}
                    \prod_{k=0}^{2s-1}\hspace{-0.08truecm}a^{(s)}
                    (\theta_j\hspace{-0.04truecm}+\hspace{-0.04truecm}
                    (s\hspace{-0.04truecm}-\hspace{-0.04truecm}\frac{1}{2}\hspace{-0.04truecm}-\hspace{-0.04truecm}k)\eta)
                    \frac{Q_1(\theta_j\hspace{-0.04truecm}+\hspace{-0.04truecm}
                    (s\hspace{-0.04truecm}-\hspace{-0.04truecm}\frac{3}{2}\hspace{-0.04truecm}-\hspace{-0.04truecm}k)\eta)}
                    {Q_2(\theta_j\hspace{-0.04truecm}+\hspace{-0.04truecm}
                    (s\hspace{-0.04truecm}-\hspace{-0.04truecm}\frac{1}{2}\hspace{-0.04truecm}-\hspace{-0.04truecm}k)\eta)},\\[6pt]
\hspace{-0.8truecm}\L^{(s-\frac{1}{2},s)}
                     (\theta_j\hspace{-0.04truecm}+\hspace{-0.04truecm}\frac{\eta}{2})\hspace{-0.28truecm}&=&\hspace{-0.32truecm}
                     \frac{Q(\theta_j\hspace{-0.04truecm}+\hspace{-0.04truecm}
                     (\frac{1}{2}\hspace{-0.04truecm}-\hspace{-0.04truecm}s)\eta)}
                     {Q(\theta_j\hspace{-0.04truecm}+\hspace{-0.04truecm}
                     (s\hspace{-0.04truecm}-\hspace{-0.04truecm}\frac{1}{2})\eta)}\hspace{-0.08truecm}
                    \prod_{k=0}^{2s-2}\hspace{-0.08truecm}a^{(s)}(\theta_j\hspace{-0.04truecm}+\hspace{-0.04truecm}
                    (s\hspace{-0.04truecm}-\hspace{-0.04truecm}\frac{1}{2}\hspace{-0.04truecm}-\hspace{-0.04truecm}k)\eta)
                    \frac{Q_1(\theta_j\hspace{-0.04truecm}+\hspace{-0.04truecm}
                    (s\hspace{-0.04truecm}-\hspace{-0.04truecm}\frac{3}{2}\hspace{-0.04truecm}-\hspace{-0.04truecm}k)\eta)}
                    {Q_2(\theta_j\hspace{-0.04truecm}+\hspace{-0.04truecm}
                    (s\hspace{-0.04truecm}-\hspace{-0.04truecm}\frac{1}{2}\hspace{-0.04truecm}-\hspace{-0.04truecm}k)\eta)},\\[6pt]
\hspace{-0.8truecm}\L^{(\frac{1}{2},s)}
                     (\theta_j\hspace{-0.04truecm}-(\hspace{-0.04truecm}\frac{1}{2}\hspace{-0.04truecm}+\hspace{-0.04truecm}s)\eta)
                     \hspace{-0.28truecm}&=&\hspace{-0.28truecm}d^{(s)}
                     (\theta_j\hspace{-0.04truecm}-\hspace{-0.04truecm}(\frac{1}{2}\hspace{-0.04truecm}+\hspace{-0.04truecm}s)\eta)
                     \frac{Q(\theta_j\hspace{-0.04truecm}+\hspace{-0.04truecm}(\frac{1}{2}-s)\eta)
                     Q_2(\theta_j\hspace{-0.04truecm}+\hspace{-0.04truecm}(\frac{1}{2}-s)\eta)}
                     {Q(\theta_j\hspace{-0.04truecm}-\hspace{-0.04truecm}(\frac{1}{2}+s)\eta)
                     Q_1(\theta_j\hspace{-0.04truecm}-\hspace{-0.04truecm}(\frac{1}{2}+s)\eta)}.
\eea The above equations give rise to
\bea
\hspace{-0.8truecm}\L^{(s,s)}(\theta_j)\L^{(\frac{1}{2},s)}
                 (\theta_j\hspace{-0.04truecm}-\hspace{-0.04truecm}(\frac{1}{2}\hspace{-0.04truecm}+\hspace{-0.04truecm}s)\eta)
&=&a^{(s)}(\theta_j\hspace{-0.04truecm}+\hspace{-0.04truecm}
                  (\frac{1}{2}\hspace{-0.04truecm}-\hspace{-0.04truecm}s)\eta)
                  d^{(s)}(\theta_j\hspace{-0.04truecm}-\hspace{-0.04truecm}
                    (\frac{1}{2}\hspace{-0.04truecm}+\hspace{-0.04truecm}s)\eta)
                  \frac{Q(\theta_j\hspace{-0.04truecm}+\hspace{-0.04truecm}
                  (\frac{1}{2}\hspace{-0.04truecm}-\hspace{-0.04truecm}s)\eta)}
                   {Q(\theta_j\hspace{-0.04truecm}+\hspace{-0.04truecm}
                  (s\hspace{-0.04truecm}-\hspace{-0.04truecm}\frac{1}{2})\eta)}\no\\[6pt]
&&\quad\times\prod_{k=0}^{2s-2}a^{(s)}(\theta_j\hspace{-0.04truecm}+\hspace{-0.04truecm}
                    (s\hspace{-0.04truecm}-\hspace{-0.04truecm}\frac{1}{2}\hspace{-0.04truecm}-\hspace{-0.04truecm}k)\eta)
                    \frac{Q_1(\theta_j\hspace{-0.04truecm}+\hspace{-0.04truecm}
                    (s\hspace{-0.04truecm}-\hspace{-0.04truecm}\frac{3}{2}\hspace{-0.04truecm}-\hspace{-0.04truecm}k)\eta)}
                    {Q_2(\theta_j\hspace{-0.04truecm}+\hspace{-0.04truecm}
                    (s\hspace{-0.04truecm}-\hspace{-0.04truecm}\frac{1}{2}\hspace{-0.04truecm}-\hspace{-0.04truecm}k)\eta)}\no\\[6pt]
&=&\d^{(s)}(\theta_j\hspace{-0.04truecm}+\hspace{-0.04truecm}
(\frac{1}{2}\hspace{-0.04truecm}-\hspace{-0.04truecm}s)\eta)\,
\L^{(s-\frac{1}{2},s)}(\theta_j\hspace{-0.04truecm}+\hspace{-0.04truecm}\frac{\eta}{2}),\, j=1,\ldots,N, \eea
indicating that
the $T-Q$ relation (\ref{T-Q-ansatz-1}) indeed satisfies the very
functional identities (\ref{Eigenvalue-identity}). From the explicit
expression (\ref{T-Q-ansatz-1}) one may find that there
might be some apparent simple poles at the following points: \bea
\l_j,\,-\l_j-\eta,\quad \mu_k,\,-\mu_k-\eta,\quad
j=1,\ldots,2sN+m-2M,\quad k=1,\ldots,2M.\label{Simple-poles} \eea As
required by the regularity of the transfer matrix, the residues of
$\L^{(\frac{1}{2},s)}(u)$ (\ref{T-Q-ansatz-1}) at these points must vanish,
which leads to the following BAEs \bea
&&a^{(s)}(\l_j)Q(\l_j-\eta)Q_1(\l_j)Q_1(\l_j-\eta) +d^{(s)}(\l_j)Q(\l_j+\eta)Q_2(\l_j)Q_2(\l_j+\eta)\no\\[2pt]
&&\quad\quad \quad\quad +c\,(\l_j(\l_j+\eta))^{m+1}\,F^{(s)}(\l_j)=0,\quad j=1,\ldots,2sN+m-2M,\label{BAE-s-1}\\[2pt]
&&d^{(s)}(\mu_k)Q(\mu_k+\eta)Q_2(\mu_k)Q_2(\mu_k+\eta)+c\,(\mu_k(\mu_k+\eta))^{m+1}\,F^{(s)}(\mu_k)=0.\,\,\no\\[2pt]
&&\quad\quad \quad\quad k=1,\ldots,2M.\label{BAE-s-2}
\eea

Finally we conclude that the $T-Q$ relation (\ref{T-Q-ansatz-1})
indeed satisfies (\ref{Eigenvalue-hier})-(\ref{Eigenvalue-Anal}) as
it is required if the $2sN+m$ parameters
$\{\l_j|j=1,\ldots,2sN+m-2M\}$ and $\{\mu_j|j=1,\ldots,2M\}$ satisfy
the associated BAEs (\ref{BAE-s-1})-(\ref{BAE-s-2}). Thus the
$\L^{(\frac{1}{2},s)}(u)$ given by (\ref{T-Q-ansatz-1}) becomes the
eigenvalue of the transfer matrix $t^{(\frac{1}{2},s)}(u)$ given by
(\ref{Fused-transfer}). With the help of the recursive relation
(\ref{Eigenvalue-hier}), we can obtain the inhomogeneous $T-Q$
equations for all the other $\L^{(j,s)}(u)$ from the fundamental one
$\L^{(\frac{1}{2},s)}(u)$.

The results of isotropic spin-$\frac12$
chains \cite{cao1,Nep13-1,Jia13}  suggest that fixed $m$ and $M$
can give a complete set of eigenvalues of the transfer matrix. In Appendix A, we prove that each solution of
(\ref{Eigenvalue-hier})-(\ref{Eigenvalue-Anal})  can be
parameterized by the  inhomogeneous $T-Q$ relation
with fixed $m$ and $M$. In such a sense, different $m$ and $M$ just give different
parameterizations of $\L(u)$ but not new solutions of $\L(u)$ \footnote{In fact, there are many ways to parameterize a polynomial function, e.g, with its zeros or with its coefficients. $T-Q$ relation is a convenient one but not the unique one to characterize the eigenvalues of the transfer matrix. Especially, with a nonzero off-diagonal term, there are more freedoms to construct  inhomogeneous $T-Q$ relations obeying the functional relations.}. Here we list some
special forms of the $T-Q$  relations for particular choices of $m$
and $M$.
\begin{itemize}
\item {\bf The case of $Q_1(u)=Q_2(u)=1$.}   In this case, $M=0$ and one can always choose $m=0$ such that the number of the Bethe parameters
$\{\l_j\}$ takes the minimal value $2sN$. The resulting $T-Q$ relation reads
\bea
 \L^{(\frac{1}{2},s)}(u)&=&a^{(s)}(u)\frac{Q(u-\eta)}{Q(u)}
                           +d^{(s)}(u)\frac{Q(u+\eta)}{Q(u)}+c\,u(u+\eta)\frac{F^{(s)}(u)}{Q(u)},\label{T-Q-ansatz-2}
\eea where the functions $a^{(s)}(u)$, $d^{(s)}(u)$, $F^{(s)}(u)$ and the constant $c$ are given by (\ref{a-function})-(\ref{c-constant}) respectively and the
associated $Q(u)$ function is
\bea
Q(u)&=&\prod_{j=1}^{2sN}(u-\l_j)(u+\l_j+\eta).\label{Q-Appendix}
\eea The $2sN$ parameters $\{\l_j\}$ satisfy the resulting BAEs
\bea
&&a^{(s)}(\l_j)Q(\l_j-\eta) +d^{(s)}(\l_j)Q(\l_j+\eta)+c\,\l_j(\l_j+\eta)\,F^{(s)}(\l_j)=0,\no\\
&&\qquad\qquad j=1,\ldots,2sN.\label{BAE-s-3}
\eea

\item {\bf The case of $Q(u)=1$.} The minimal value of $m$ in this case does depend on the parity of $2sN$. If $2sN$ is even,
one can choose $m=0$ such that the number of the Bethe parameters
$\{\mu_j\}$ is $2sN$. The resulting $T-Q$ relation becomes \bea
\hspace{-0.8truecm}
\L^{(\frac{1}{2},s)}(u)&=&a^{(s)}(u)\frac{Q_1(u-\eta)}{Q_2(u)}
                           +d^{(s)}(u)\frac{Q_2(u+\eta)}{Q_1(u)}
                           +c\,u(u+\eta)\frac{F^{(s)}(u)}{Q_1(u)\,Q_2(u)},\label{T-Q-ansatz-3}
\eea where \bea
Q_1(u)&=&\prod_{j=1}^{M_1}(u-\mu_j)=Q_2(-u-\eta),\quad M_1=2sN.
\label{Q1-2} \eea The resulting BAEs read \bea
d^{(s)}(\mu_j)Q_2(\mu_j)Q_2(\mu_j+\eta)+c\,\mu_j(\mu_j+\eta)\,F^{(s)}(\mu_j)=0,\,\,
j=1,\ldots,M_1.\label{BAE-s-4} \eea On the other hand, if $2sN$ is odd, the minimal $m$ becomes $m=1$ and the
corresponding number of the Bethe parameters $\{\mu_j\}$ is $2sN+1$.
The associated $T-Q$ relation is \bea \hspace{-0.8truecm}
\L^{(\frac{1}{2},s)}(u)&=&\hspace{-0.04truecm}a^{(s)}(u)\frac{Q_1(u\hspace{-0.04truecm}-\hspace{-0.04truecm}\eta)}{Q_2(u)}
                           +d^{(s)}(u)\frac{Q_2(u\hspace{-0.04truecm}+\hspace{-0.04truecm}\eta)}{Q_1(u)}
                           +c\,u^2(u\hspace{-0.04truecm}+\hspace{-0.04truecm}\eta)^2
                           \frac{F^{(s)}(u)}{Q_1(u)\,Q_2(u)},\label{T-Q-ansatz-4}
\eea where $Q_1(u)$ is still given by (\ref{Q1-2}) but with $M_1=2sN+1$ and  the resulting BAEs now become
\bea
d^{(s)}(\mu_j)Q_2(\mu_j)Q_2(\mu_j+\eta)+c\,\mu_j^2(\mu_j+\eta)^2\,F^{(s)}(\mu_j)=0,\,\, j=1,\ldots,2sN+1.\label{BAE-s-5}
\eea
\end{itemize}

It should be remarked that there also exist other choices for the functions $a^{(s)}(u)$, $d^{(s)}(u)$ and the constant $c$. For  $\{\e_i=\pm 1|i=1,2,3\}$\footnote{Such discrete variables were used to construct the $T-Q$ relation for
spin-$\frac{1}{2}$ XXZ open chain \cite{Yan06}.}, let us introduce
\bea
a^{(s)}(u;\e_1,\e_2,\e_3)&=&\e_1\frac{2u+2\eta}{2u+\eta}(\sqrt{1+\a_+^2}\,u+\e_2 p_+)(\sqrt{1+\a_-^2}\,u+\e_3 p_-) \no\\[2pt]
          &&\quad \times \prod_{l=1}^N(u-\theta_l+(\frac{1}{2}+s)\eta)(u+\theta_l+(\frac{1}{2}+s)\eta),\label{a-2-function}\\[2pt]
d^{(s)}(u;;\e_1,\e_2,\e_3)&=&a^{(s)}(-u-\eta;\e_1,\e_2,\e_3),\no\\[2pt]
c(\e_1,\e_2,\e_3)&=&2(\a_-\a_+-1-\e_1\sqrt{(1+\a^2_-)(1+\a^2_+)}).\label{c-2-constant}
\eea Similarly as in \cite{Yan06}, the three discrete variables $\{\e_i\}$ are required to obey  the following relation
\bea
\e_1\e_2\e_3=1.\label{Discrete-Constraint}
\eea Alternatively, let us make the following $T-Q$ ansatz for the eigenvalue $\Lambda^{(\frac{1}{2},s)}(u)$
\bea
 \L^{(\frac{1}{2},s)}(u)&=&a^{(s)}(u;\e_1,\e_2,\e_3)\frac{Q(u-\eta)\,Q_1(u-\eta)}{Q(u)\,Q_2(u)}
                           +d^{(s)}(u;\e_1,\e_2,\e_3)\frac{Q(u+\eta)\,Q_2(u+\eta)}{Q(u)\,Q_1(u)}\no\\[6pt]
 &&+c(\e_1,\e_2,\e_3)\,u(u+\eta)\frac{F^{(s)}(u)}{Q(u)\,Q_1(u)\,Q_2(u)},\label{T-Q-ansatz-2-1}
\eea where $F^{(s)}(u)$ is given by  (\ref{F-function}) and the $Q$-functions are given by (\ref{Q-function})-(\ref{Q2-function}) with $m=0$.
It is easy to check that the alternative $T-Q$ relation (\ref{T-Q-ansatz-2-1})
indeed satisfies (\ref{Eigenvalue-hier})-(\ref{Eigenvalue-Anal}) if the $2sN$ parameters
$\{\l_j|j=1,\ldots,2sN-2M\}$ and $\{\mu_j|j=1,\ldots,2M\}$ satisfy
the similar BAEs  as (\ref{BAE-s-1})-(\ref{BAE-s-2}) but with the functions $a^{(s)}(u)$,  $d^{(s)}(u)$ and the constant $c$ replaced by
$a^{(s)}(u;\e_1,\e_2,\e_3)$,  $d^{(s)}(u;\e_1,\e_2,\e_3)$ and $c(\e_1,\e_2,\e_3)$, respectively.  Each choice of $\{\e_i\}$ satisfying the constraint (\ref{Discrete-Constraint}) can give the complete
set of the spectrum. Moreover, if the boundary parameters satisfy the constraint $\a_-=-\a_+$, which corresponds to that the two $K^{\pm}(u)$ can
be diagonalized simultaneously, the algebraic Bethe ansatz method can be applied \cite{matins}. In this particular
case one can choose $\e_1=-1$, $\e_2\e_3=-1$ and therefore $c=0$. The corresponding $T-Q$ ansatz (\ref{T-Q-ansatz-2-1}), under the similar analysis as that
in \cite{cao1}, is naturally reduced to the conventional one \cite{matins}
obtained by the algebraic Bethe ansatz.

\subsection{Spin-$1$ case}
In this subsection we illustrate the completeness of the Bethe
ansatz solutions obtained in the previous subsection. For the case
of $s=\frac{1}{2}$, which corresponds to the spin-$\frac{1}{2}$ XXX
spin chain and the corresponding transfer matrix is
$t^{(\frac{1}{2},\frac{1}{2})}(u)$, our result is reduced to that
obtained in \cite{cao1}. The completeness of the Bethe ansatz
solution was already studied in \cite{cao1,Nep13-1,Jia13}. Here we
provide numerical evidence for the $s=1$ case, which corresponds to
the isotropic Fateev-Zamolodchikov (or Takhtajan-Babujian) model
\cite{zf,spinsXXX} with general non-diagonal boundary terms. In
terms of the basis $\{|l\rangle|l=1,0,-1\}$ given by \bea
&&|1\rangle =|\frac{1}{2}\rangle\otimes|\frac{1}{2}\rangle,\no\\[6pt]
&&|0\rangle =\frac{1}{\sqrt{2}}\lt(|\frac{1}{2}\rangle\otimes|-\frac{1}{2}\rangle+|-\frac{1}{2}\rangle\otimes|\frac{1}{2}\rangle\rt),\no\\[6pt]
&&|-1\rangle =|-\frac{1}{2}\rangle\otimes|-\frac{1}{2}\rangle,\no
\eea
 the corresponding spin-$(1,1)$ $R$-matrix $R^{(1,1)}(u)$ defined in (\ref{fusedR010013}) is

\bea
 R^{(1,1)}(u)=\left(\begin{array}{r|r|r}{\begin{array}{rrr}c(u)&&\\&b(u)&\\&&d(u)\end{array}}
           &{\begin{array}{lll}&&\\e(u)&{\,\,\,\,\,\,\,\,\,\,}&{\,\,\,\,\,\,\,\,\,\,}\\{\,\,\,\,\,\,}&g(u)&{\,\,\,\,\,\,}\end{array}}
           &{\begin{array}{lll}&&\\&&\\f(u)&{\,\,\,\,\,\,\,\,\,\,}&{\,\,\,\,\,\,\,\,\,\,}\end{array}}\\[12pt]
 \hline {\begin{array}{rrr}&e(u)&\\&&g(u)\\&&\end{array}}&
           {\begin{array}{ccc}b(u)&&\\&a(u)&\\&&b(u)\end{array}}
           &{\begin{array}{lll}&&\\g(u)&{\,\,\,\,\,\,\,\,\,\,}&{\,\,\,\,\,\,\,\,\,\,}\\&e(u)&{\,\,\,\,\,\,\,\,\,\,}\end{array}}\\[12pt]
 \hline {\begin{array}{ccc}&&f(u)\\&&\\&&\end{array}}
           &{\begin{array}{ccc}&g(u)&\\&&e(u)\\&&\end{array}}
           &{\begin{array}{ccc}d(u)&&\\&b(u)&\\&&c(u)\end{array}} \end{array}\right),\label{R-matrix-11}
\eea

\noindent where the non-vanishing entries are
\bea
&&a(u)=u(u+\eta)+2\eta^2,\,\, b(u)=u(u+\eta),\,\,c(u)=(u+\eta)(u+2\eta),\no\\
&&d(u)=u(u-\eta),\,\,e(u)=2\eta(u+\eta),\,\, f(u)=2\eta^2,\,\,g(u)=2u\eta.\label{R-matrix-element}
\eea

The spin-$1$ $K$-matrix defined by
(\ref{fusedKmatrix0103}), in terms of the basis
$\{|l\rangle|l=1,0,-1\}$, is given by

\bea
K^{-(1)}(u)=(2u+\eta)\left(\begin{array}{ccc}x_1(u)&y_4(u)&y_6(u)\\
y_4(u)&x_2(u)&y_5(u)\\
y_6(u)&y_5(u)&x_3(u)\end{array}\right), \label{K-matrix-1-1}
\eea

\noindent where the matrix elements are

\bea
x_1(u)&=&(p_-+u+\frac{\eta}{2})\,(p_-+u-\frac{\eta}{2})+\frac{\a_-^2}{2}\,\eta\,(u-\frac{\eta}{2}),\no\\[6pt]
x_2(u)&=&(p_-+u-\frac{\eta}{2})\,(p_--u+\frac{\eta}{2})+\a_-^2\,(u+\frac{\eta}{2})\,(u-\frac{\eta}{2}),\no\\[6pt]
x_3(u)&=&(p_--u-\frac{\eta}{2})\,(p_--u+\frac{\eta}{2})+\frac{\a_-^2}{2}\,\eta\,(u-\frac{\eta}{2}),\no\\[6pt]
y_4(u)&=&\sqrt{2}\,\a_-\,u\,(p_-+u-\frac{\eta}{2}),\no\\[6pt]
y_5(u)&=&\sqrt{2}\,\a_-\,u\,(p_--u+\frac{\eta}{2}),\no\\[6pt]
y_6(u)&=&\a_-^2\,u\,(u-\frac{\eta}{2}).\label{K-matrix-1-2}
\eea

\noindent The dual spin-$1$ $K$-matrix $K^{+(1)}(u)$ can be given by the above $K$-matrix through the correspondence (\ref{Correspond}).

The eigenvalue $\L^{(\frac{1}{2},1)}(u)$ in the homogeneous limit (i.e.,
$\theta_j\rightarrow 0$) reads
\bea
 \L^{(\frac{1}{2},1)}(u)&=&a^{(1)}(u)\frac{Q(u-\eta)\,Q_1(u-\eta)}{Q(u)\,Q_2(u)}
                           +d^{(1)}(u)\frac{Q(u+\eta)\,Q_2(u+\eta)}{Q(u)\,Q_1(u)}\no\\[6pt]
 &&+c\,u(u+\eta)\frac{F^{(1)}(u)}{Q(u)\,Q_1(u)\,Q_2(u)},\label{T-Q-ansatz-5}
\eea where we have chosen $m=0$ and the functions $a^{(1)}(u)$, $d^{(1)}(u)$, $F^{(1)}(u)$ are given by
\bea
a^{(1)}(u)&=&\frac{2u+2\eta}{2u+\eta}(\sqrt{1+\a_+^2}\,u+p_+)(\sqrt{1+\a_-^2}\,u+p_-)(u+\frac{3\eta}{2})^{2N},\label{a-function-1}\\[6pt]
d^{(1)}(u)&=&a^{(1)}(-u-\eta),\label{d-function-1}\\[6pt]
F^{(1)}(u)&=&(u-\frac{\eta}{2})^{2N}(u+\frac{\eta}{2})^{2N}(u+\frac{3\eta}{2})^{2N}.\label{F-function-1}
\eea The constant $c$ is given by (\ref{c-constant}). The three
$Q$-functions are parameterized by $2N$ parameters
$\{\l_j|j=1,\ldots,2N-2M\}$ and $\{\mu_j|j=1,\ldots,2M\}$ (with $M$
a non-negative integer) as \bea
Q(u)&=&\prod_{j=1}^{2N-2M}(u-\l_j)(u+\l_j+\eta)=Q(-u-\eta),\label{Q-1-function}\\[2pt]
Q_1(u)&=&\prod_{j=1}^{2M}(u-\mu_j)=Q_2(-u-\eta),\label{Q1-1-function}\\[2pt]
Q_2(u)&=&\prod_{j=1}^{2M}(u+\mu_j+\eta)=Q_1(-u-\eta).\label{Q2-1-function}
\eea
The $2N$ parameters $\{\l_j|j=1,\ldots,2N-2M\}$ and $\{\mu_j|j=1,\ldots,2M\}$ satisfy the following
BAEs
\bea
&&a^{(1)}(\l_j)Q(\l_j-\eta)Q_1(\l_j)Q_1(\l_j-\eta) +d^{(1)}(\l_j)Q(\l_j+\eta)Q_2(\l_j)Q_2(\l_j+\eta)\no\\[2pt]
&&\quad\quad \quad\quad +c\,\l_j(\l_j+\eta)\,F^{(1)}(\l_j)=0,\quad j=1,\ldots,2N-2M,\label{BAE-1-1}\\[2pt]
&&d^{(1)}(\mu_k)Q(\mu_k+\eta)Q_2(\mu_k)Q_2(\mu_k+\eta)+c\,\mu_k(\mu_k+\eta)\,F^{(1)}(\mu_k)=0,\,\,\no\\[2pt]
&&\quad\quad \quad\quad k=1,\ldots,2M.\label{BAE-1-2} \eea
The eigenvalue $\L^{(1,1)}(u)$ can be constructed
from the fundamental one $\L^{(\frac{1}{2},1)}(u)$ given by
(\ref{T-Q-ansatz-5})-(\ref{F-function-1}) by using the relation
(\ref{Eigenvalue-hier})  as follows \bea \L^{(1,1)}(u)&=&
\L^{(\frac{1}{2},1)}(u+\frac{\eta}{2})\,\L^{(\frac{1}{2},1)}(u-\frac{\eta}{2})-\d^{(1)}(u+\frac{\eta}{2}).\label{Eigen-1-1}
\eea The Hamiltonian of the spin-1 XXX open chain with the generic
non-diagonal boundary terms is given by \bea H&=&\partial_u\lt\{\ln
u(u+\eta)\,t^{(1,1)}(u)\rt\}\lt|_{u=0}\rt. \nonumber \\&=&
\frac{1}{\eta^2}\sum_{j=1}^{N-1} \left[\vec S_{j} \cdot \vec S_{j+1} -(\vec S_{j} \cdot \vec S_{j+1})^2 \right]  \no \\
&& + \frac{1}{p_-^2-\frac{1}{4}(1+\alpha_-^2)\eta^2}\left[
2p_-\alpha_-S_{1}^x+2p_-S_{1}^z +\frac12
(\alpha_-^2\eta-2\eta) (S_{1}^z )^2  \right. \no \\
&& \qquad \left. -\frac12 \alpha_-^2\eta[(S_{1}^x )^2-(S_{1}^y )^2] -\alpha_-\eta[S_{1}^zS_{1}^x+S_{1}^xS_{1}^z]\right] \no \\
&& +\frac{1}{(3p_+^2-\frac{3}{4}(1+\alpha_+^2)\eta^2)\eta^2}\left[ 6
p_+ \alpha_+\eta S_{N}^x -6 p_+ \eta S_{N}^z
 \right. \no \\[6pt]
&&\qquad  \left. +3\alpha_+\eta^2 [S_{N}^xS_{N}^z + S_{N}^zS_{N}^x] -(2p_+^2-\frac{3}{2}(1-\alpha_+^2)\eta^2) (S_{N}^x )^2 \right. \no \\[6pt]
&&\qquad  \left. -(2p_+^2-\frac{3}{2}(1+\alpha_+^2)\eta^2) (S_{N}^y )^2 -(2p_+^2+\frac{3}{2}(1-\alpha_+^2)\eta^2) (S_{N}^z )^2
\right] \no \\
&& +\frac{\eta(1+\alpha_+^2)}
{3p_+^2-\frac{3}{4}(1+\alpha_+^2)\eta^2} +\frac{\eta}
{p_-^2-\frac{1}{4}(1+\alpha_-^2)\eta^2}
+3N\frac1{\eta^2}+\frac{4}{\eta}. \label{Ham-1-1} \eea The
eigenvalues of the Hamiltonian (\ref{Ham-1-1}) thus read \bea
E&=&\sum_{j=1}^{2N-2M}\frac{4\eta}{(\l_j+\frac{3\eta}{2})(\l_j-\frac{\eta}{2})}-\sum_{k=1}^{2M}
\frac{4(\mu_k+\eta)}{(\mu_k+\frac{\eta}{2})(\mu_k+\frac{3\eta}{2})}+E_0,\label{E-1-1}\\[6pt]
E_0&=&\frac{1}{\eta}\lt\{3N+\frac{8}{3}+\frac{2\sqrt{1+\a_+^2}\,p_+\eta}{p_+^2-\frac{\eta^2}{4}(1+\a_+^2)}+
\frac{2\sqrt{1+\a_-^2}\,p_-\eta}{p_-^2-\frac{\eta^2}{4}(1+\a_-^2)}\rt\}.
\eea
\begin{table}
\caption{Solutions of the BAEs (\ref{BAE-1-1})-(\ref{BAE-1-2}) for $N=2$,
$M=0$, $\eta=1$, $p_+=0.1$, $p_-=0.2$, $\alpha_+=0.3$ and
$\alpha_-=0.4$. $n$ indicates the number of the energy levels and $E_n$ is the corresponding eigenenergy. The energy $E_n$ calculated from the Bethe roots is exactly the same to that from the
exact diagonalization of the Hamiltonian (\ref{Ham-1-1}).}{\footnotesize
\begin{center}
\begin{tabular}{cccc|c|c} \hline\hline
$\lambda_1$ & $\lambda_2$ & $\lambda_3$ & $\lambda_4$ & $E_n$ & $n$
\\ \hline
$0.02022$ & $0.15565-0.56301i$ & $0.15565+0.56301i$ & $1.28344$ & $-2.82985$ & $1$ \\
$0.01436-0.14539i$ & $0.01436+0.14539i$ & $1.02580-0.23475i$ & $1.02580+0.23475i$ & $0.74454$ & $2$ \\
$0.00579-0.12153i$ & $0.00579+0.12153i$ & $0.95719-0.17885i$ & $0.95719+0.17885i$ & $1.84509$ & $3$ \\
$-0.50000+0.46805i$ & $-0.09323$ & $0.93690$ & $1.18821$ & $3.99277$ & $4$ \\
$0.06934-0.91728i$ & $0.06934+0.91728i$ & $1.06778-0.60960i$ & $1.06778+0.60960i$ & $4.36850$ & $5$ \\
$-0.50000+0.16632i$ & $-0.18832$ & $0.82026$ & $1.19558$ & $5.34163$ & $6$ \\
$-0.09561$ & $0.89614$ & $1.31281-0.54820i$ & $1.31281+0.54820i$ & $7.59257$ & $7$ \\
$-0.25439$ & $0.03756$ & $0.73530-0.09425i$ & $0.73530+0.09425i$ & $9.12855$ & $8$ \\
$-0.18554$ & $0.81124$ & $1.29199-0.51363i$ & $1.29199+0.51363i$ & $9.43905$ & $9$ \\
\hline\hline \end{tabular}
\end{center}}
\end{table}
\begin{table}
\caption{Solutions of the BAEs (\ref{BAE-1-1})-(\ref{BAE-1-2}) for $N=2$,
$M=2$, $\eta=1$, $p_+=0.1$, $p_-=0.2$, $\alpha_+=0.3$ and
$\alpha_-=0.4$. $n$ indicates the number of the energy levels and $E_n$ is the corresponding eigenenergy. The energy $E_n$ calculated from the Bethe roots is exactly the same to that from the
exact diagonalization of the Hamiltonian (\ref{Ham-1-1}).
}{\footnotesize
\begin{center}
\begin{tabular}{cccc|c|c} \hline\hline
$\mu_1$ & $\mu_2$ & $\mu_3$ & $\mu_4$ & $E_n$ & $n$ \\ \hline
$-2.01449$ & $-1.03956$ & $-0.20728-0.16066i$ & $-0.20728+0.16066i$ & $-2.82985$ & $1$ \\
$-1.02843$ & $-0.99277$ & $0.08448$ & $9.58424$ & $0.74454$ & $2$ \\
$-1.24529$ & $-0.75827$ & $0.29385$ & $6.34119$ & $1.84509$ & $3$ \\
$-0.90627$ & $-0.50220-0.45722i$ & $-0.50220+0.45722i$ & $8.15074$ & $3.99277$ & $4$ \\
$-6.63560$ & $-1.41141$ & $-0.58859$ & $0.74744$ & $4.36850$ & $5$ \\
$-0.80847$ & $-0.50016-0.16539i$ & $-0.50016+0.16539i$ & $6.62545$ & $5.34163$ & $6$ \\
$-4.94473$ & $-0.90422$ & $2.81138$ & $35.70597$ & $7.59257$ & $7$ \\
$-4.56687$ & $-0.88866$ & $-0.82335$ & $0.40258$ & $9.12855$ & $8$ \\
$-4.49000$ & $-0.81430$ & $2.38972$ & $32.53964$ & $9.43905$ & $9$ \\
\hline\hline \end{tabular}
\end{center}}
\end{table}
\begin{figure}[ht]
\begin{center}
\includegraphics[width=10cm,height=7cm]{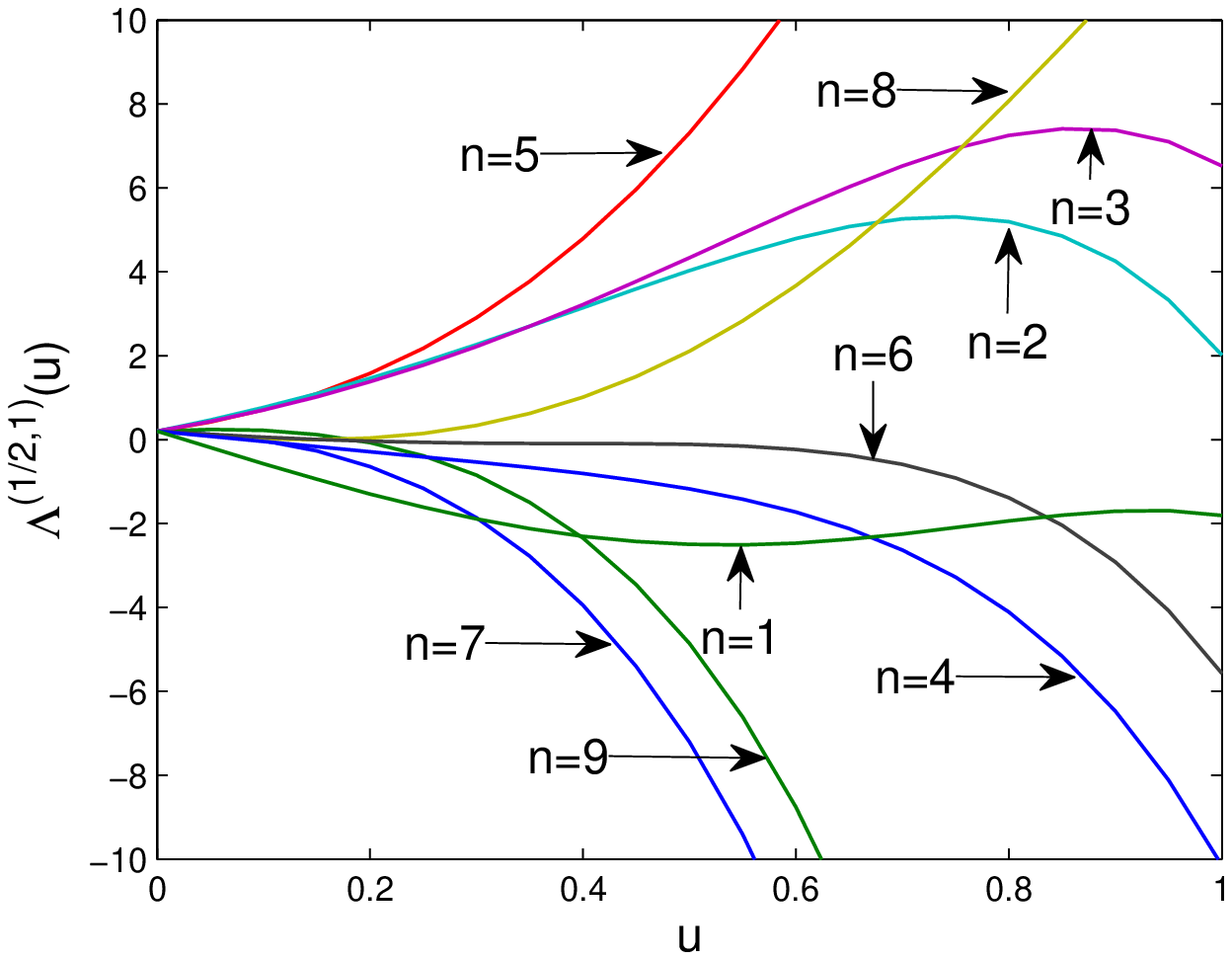}\\
\caption{$\Lambda^{(\frac12,1)}(u)$ vs. $u$ calculated from both the
$T-Q$ relations and exact diagonalization of $t^{(\frac12,1)}(u)$
for $N=2$, $\eta=1$, $p_+=0.1$, $p_-=0.2$, $\alpha_+=0.3$,
$\alpha_-=0.4$. The correspondence is indicated by the number $n=1$
to $9$.} \label{N2Lambda3}\end{center}
\end{figure}
Numerical solutions of the BAEs and exact diagonalizations of the transfer matrix $t^{(\frac12,1)}(u)$ and the
Hamiltonian (\ref{Ham-1-1}) are performed for the case of $N=2$ and randomly
choosing of boundary parameters. The results are listed in Table 1 for
$M=0$ and Table 2 for $M=2$, respectively. The eigenvalues of the Hamiltonian obtained by
solving the BAEs are exactly the same to those obtained by the exact
diagonalization of the Hamiltonian. The eigenvalues
$\L^{(\frac12,1)}(u)$ of the transfer matrix $t^{(\frac12,1)}(u)$ are shown in
Figure 1. Again, the curves of $\L^{(\frac12,1)}(u)$ calculated from
the BAEs and the $T-Q$ relations coincide exactly with those from
the exact diagonalization of the transfer matrix
$t^{(\frac12,1)}(u)$. The numerical results strongly suggest that a
fixed $M$ is enough to give the complete spectrum of the transfer
matrix.

\section{Conclusions} \setcounter{equation}{0}

The spin-$s$ XXX chain with the generic non-diagonal boundary terms
specified by the most general non-diagonal $K$-matrices given by
(\ref{fusedKmatrix0103})-(\ref{K-0102}) has been studied by the
off-diagonal Bethe anstz method.  Based on the intrinsic properties
of the fused $R$-matrices and $K$-matrices, we obtain the closed
operator identities (\ref{Operator-identity-1}) of the fundamental
transfer matrix $t^{(\frac{1}{2},s)}(u)$. These identities, together
with other properties
(\ref{Operator-Int-1})-(\ref{Operator-crossing}), allow us to
construct an off-diagonal (or inhomogeneous) $T-Q$ equation
(\ref{T-Q-ansatz-1}) of the eigenvalues of the transfer matrix and
the associated BAEs (\ref{BAE-s-1})-(\ref{BAE-s-2}).  It should be
emphasized that there are a variety  of forms of the inhomogeneous
$T-Q$ relations such as (\ref{T-Q-ansatz-2}), (\ref{T-Q-ansatz-3})
and (\ref{T-Q-ansatz-4}). Each of them should give the complete spectrum of the transfer matrix.
Taking the spin-$1$ XXX chain as an example, we give the numerical evidence
for two-site $s=1$ case. We note that the method developed in the present
paper can also be applied to the integrable models with cyclic
representations such as the lattice sine-Gordon model, the $\tau_2$
model, the relativistic Toda chain and the chiral Potts model with generic integrable boundary
conditions.

For the spin-$\frac{1}{2}$ case, the Bethe states corresponding to the $T-Q$ relation
(\ref{T-Q-ansatz-2}) were constructed in \cite{Cao-14-Bethe-state} (also conjectured in \cite{sigma})
with the helps of the SoV basis proposed in \cite{Nic12}. It is interesting that the resulting Bethe states directly induces the homogeneous limits of the SoV states constructed in \cite{Nic12}.
Following the similar procedure, the eigenstates for the spin-$s$ open chains might be constructed with a similar basis proposed in \cite{Nicco}\footnote{Alternatively, one should take the eigenstates of an off-diagonal elements of the double-row monodromy matrix to form a basis.}.

\section*{Acknowledgments}

The financial support from  the National Natural Science Foundation
of China (Grant Nos. 11375141, 11374334, 11434013, 11425522), the
National Program for Basic Research of MOST (973 project under grant
No.2011CB921700), the State Education Ministry of China (Grant No.
20116101110017) and BCMIIS are gratefully acknowledged. Two of the
authors (W.-L. Yang and K. Shi) would like to thank IoP, CAS for the
hospitality during their visit. W.-L. Yang and Y. Wang  acknowledge M. Martins for his helpful communication.


\section*{Appendix A: Proof of the inhomogeneous $T-Q$ relation }
\setcounter{equation}{0}
\renewcommand{\theequation}{A.\arabic{equation}}

In this appendix, we  show
that each solution of  (\ref{Eigenvalue-hier})-(\ref{Eigenvalue-Anal})  can be
parameterized  in terms of the inhomogeneous $T-Q$ relation (\ref{T-Q-ansatz-1}) with two
fixed non-negative integers $m$ and $M$ (taking $m=M=0$ as an example).

Let us introduce a function $f(u)$ associated with each solution $\L(u)$ of (\ref{Eigenvalue-hier})-(\ref{Eigenvalue-Anal})
\bea
 f(u)\hspace{-0.08truecm}=\hspace{-0.08truecm}\L^{(\frac{1}{2},s)}(u)Q(u)\hspace{-0.04truecm}-\hspace{-0.04truecm}a^{(s)}(u)
      Q(u\hspace{-0.04truecm}-\hspace{-0.04truecm}\eta)
      \hspace{-0.04truecm}-\hspace{-0.04truecm}d^{(s)}(u)
      Q(u\hspace{-0.04truecm}+\hspace{-0.04truecm}\eta)
      -\hspace{-0.04truecm}c\hspace{-0.04truecm}\,u(u\hspace{-0.04truecm}+\hspace{-0.04truecm}\eta)
      F^{(s)}(u),\label{f-function-Appendix}
\eea where the functions $a^{(s)}(u)$, $d^{(s)}(u)$, $F^{(s)}(u)$, $c$ and $Q(u)$ are given by
(\ref{a-function})-(\ref{c-constant})  and (\ref{Q-Appendix}) respectively. It follows from its definition
that the function $f(u)$, as a function of $u$, is a polynomial of degree $2(2s+1)N+2$ with the crossing symmetry
\bea
f(-u-\eta)=f(u). \label{A-2}
\eea This very property implies that the function can be fixed by its values at $(2s+1)N+2$ different points.
It is  clear from the relations (\ref{Eigenvalue-Int-1}) and (\ref{Eigenvalue-Asy}) that
\bea
f(0)=0,\quad {\rm and}\,\lim_{u\rightarrow\infty} f(u)=0\times u^{2(2s+1)N+2}+\ldots.\label{A-3}
\eea This means that it is enough to completely determine $f(u)$ by fixing its values at other $(2s+1)N$ independent points.
Thanks to the fact that $Q(u)$ is also a crossing polynomial (i.e., $Q(-u-\eta)=Q(u)$) of degree $4sN$ with a known coefficient of
the term $u^{4sN}$,
for each solution $\L(u)$ of (\ref{Eigenvalue-hier})-(\ref{Eigenvalue-Anal}) one can always choose the function $Q(u)$ of form
(\ref{Q-Appendix}) such that the following equations hold:
\bea
f(\theta_j+(s-\frac{1}{2}-k)\eta)=0,\quad k=0,\ldots,2s,\,{\rm and }\quad j=1,\ldots,N.\label{A-4}
\eea Then the relations (\ref{A-2})-(\ref{A-4}) leads to $f(u)=0$ or that each solution of  (\ref{Eigenvalue-hier})-(\ref{Eigenvalue-Anal})  can be
parameterized  in terms of the inhomogeneous $T-Q$ relation (\ref{T-Q-ansatz-2}) with properly choice of the function $Q(u)$.
In fact the conditions (\ref{A-4}) are equivalent
to the following $(2s+1)N$ linear equations with respect to the values of $Q(u)$ at the $(2s+1)N$ different points $\{\theta_j+(s-\frac{1}{2}-k)\eta|k=0,\ldots,2s,\,j=1,\ldots,N\}$, namely,
\bea
B^{(j)}\,X^{(j)}=0,\quad j=1,\ldots,N, \label{A-5}
\eea
with each $(2s+1)\times (2s+1)$ matrix $B^{(j)}$ is given by
\bea
\left(\begin{array}{cccccc}\L(\theta_j\hspace{-0.08truecm}+\hspace{-0.08truecm}(s\hspace{-0.08truecm}-\hspace{-0.08truecm}\frac{1}{2})\eta)&
-a(\theta_j\hspace{-0.08truecm}+\hspace{-0.08truecm}(s\hspace{-0.08truecm}-\hspace{-0.08truecm}\frac{1}{2})\eta)&&&&\\[4pt]
-d(\theta_j\hspace{-0.08truecm}+\hspace{-0.08truecm}(s\hspace{-0.08truecm}-\hspace{-0.08truecm}\frac{3}{2})\eta)&
\L(\theta_j\hspace{-0.08truecm}+\hspace{-0.08truecm}(s\hspace{-0.08truecm}-\hspace{-0.08truecm}\frac{3}{2})\eta)&
-a(\theta_j\hspace{-0.08truecm}+\hspace{-0.08truecm}(s\hspace{-0.08truecm}-\hspace{-0.08truecm}\frac{3}{2})\eta)&&&\\[16pt]
&\ddots&\ddots&\ddots&&\\[16pt]
&&&
-d(\theta_j\hspace{-0.08truecm}-\hspace{-0.08truecm}(s\hspace{-0.08truecm}+\hspace{-0.08truecm}\frac{1}{2})\eta)&
\L(\theta_j\hspace{-0.08truecm}-\hspace{-0.08truecm}(s\hspace{-0.08truecm}+\hspace{-0.08truecm}\frac{1}{2})\eta)&
\end{array}\right),\no
\eea
and the $(2s+1)$ components vector $X^{(j)}$ is given by
\bea
\left(\begin{array}{c}Q(\theta_j+(s-\frac{1}{2})\eta)\\[4pt]
Q(\theta_j+(s-\frac{3}{2})\eta)\\[4pt]
\vdots\\[4pt]
Q(\theta_j-(s+\frac{1}{2})\eta)
\end{array}
\right).\no
\eea
The conditions that the  $(2s+1)N$ linear equations (\ref{A-5}) have non-zero solutions is that the determinant
of each matrix  $B^{(j)}$  vanishes, namely, ${\rm Det} (B^{(j)})=0$. In this case, the number of independent linear equations (\ref{A-5}) is reduced to $2sN$ and one can always fix at most the $2sN$ values\footnote{In some degenerate case, e.g., the boundary fields are parallel, (\ref{A-5}) may correspond more $Q(u)$ solutions with $M<2sN$ and $c=0$. Equivalently, one may fix $M=2sN$, $c=0$ but take some of the Bethe roots to be infinity in the reduced homogeneous $T-Q$ relation.} (up to a scaling factor) of $Q(u)$ at $2sN$ points among $\{\theta_j+(s-\frac{1}{2}-k)\eta|k=0,1,\ldots,2s,\,j=1,\ldots,N\}$, for an example,   $\{\theta_j+(s-\frac{1}{2}-k)\eta|k=1,\ldots,2s,\,j=1,\ldots,N\}$.
Direct calculation shows that the vanishing of the determinants
of $B^{(j)}$ are exactly the very identities (\ref{Eigenvalue-identity}). Therefore,  each solution $\L(u)$ of
(\ref{Eigenvalue-hier})-(\ref{Eigenvalue-Anal}) allows one to parameterize it in terms of the inhomogeneous $T-Q$ relation
(\ref{T-Q-ansatz-2}) where  $Q(u)$ can be determined either by its  values at $2sN$ different points via the equations (\ref{A-5})
or its roots: $\{\l_j|j=1,\ldots,2sN\}$ in (\ref{Q-Appendix})  via the associated BAEs (\ref{BAE-s-3}).

One can use the similar method to check that each solution of  (\ref{Eigenvalue-hier})-(\ref{Eigenvalue-Anal})  can be also
parameterized  in terms of the inhomogeneous $T-Q$ relation with other values of $m$ and $M$. In this case the degree of
the corresponding function $f(u)$ becomes $(2s+1)N+m$. Thanks to the relation (\ref{Eigenvalue-hier})-(\ref{Eigenvalue-Anal}),
besides the $(2s+1)N$ conditions (\ref{A-4}), one is always able to choose its values at some extra $m$ points
 to be zero such that the associated $T-Q$ satisfied.

\end{document}